\documentclass[11pt]{article}
\usepackage{xspace,amssymb,amsmath,helvet,graphicx}
\makeatletter
\renewcommand\@biblabel[1]{}
\makeatother
\begin{document}
\newcommand{\dee}{\,\mbox{d}}
\newcommand{\naive}{na\"{\i}ve }
\newcommand{\eg}{e.g.\xspace}
\newcommand{\ie}{i.e.\xspace}
\newcommand{\pdf}{pdf.\xspace}
\newcommand{\etc}{etc.\@\xspace}
\newcommand{\PhD}{Ph.D.\xspace}
\newcommand{\MSc}{M.Sc.\xspace}
\newcommand{\BA}{B.A.\xspace}
\newcommand{\MA}{M.A.\xspace}
\newcommand{\role}{r\^{o}le}
\newcommand{\signoff}{\hspace*{\fill} Rose Baker \today}
\newenvironment{entry}[1]%
{\begin{list}{}{\renewcommand{\makelabel}[1]{\textsf{##1:}\hfil}%
\settowidth{\labelwidth}{\textsf{#1:}}%
\setlength{\leftmargin}{\labelwidth}
\addtolength{\leftmargin}{\labelsep}
\setlength{\itemindent}{0pt}
}}%
{\end{list}}
\title{A new distribution for robust least squares}
\author{Rose Baker and Dan Jackson\\
Salford Business School, University of Salford, UK\\
MRC Biostatistical Unit, Cambridge, UK.}
\maketitle
\begin{abstract}
A new distribution is introduced, which we call the twin-t distribution. This distribution is heavy-tailed like the t distribution,
but closer to normality in the central part of the curve. Its properties are described, \eg the pdf, the distribution function, moments, and
random number generation. This distribution could have
many applications, but here we focus on its use as an aid to robustness. We
give examples of its application in robust regression and in curve fitting. Extensions
such as skew and multivariate twin-t distributions, and a twin
of the McDonald-Newey generalised t distribution are also discussed.
\end{abstract}
\section*{Keywords}
robust regression; skew distributions; t distribution; arcsinh transformation; heavy tail.
\section{Introduction}
Given that `all models are wrong, but some are useful' (Box and Draper, 1987), statisticians need
methodologies that are robust against small departures from the model
assumptions. A major source of departure from modelling assumptions is through
the data distribution having a long tail, whether there are only a few
aberrant observations (outliers) or many, as for example with financial
returns data. Hence robustness to long tails is crucial, and the t distribution is commonly used for this purpose.

We have developed a new distribution, heavy-tailed like the t distribution,
that is closer to normality in the main body of the distribution. This could
have many applications to data modelling. Here we describe the application of our new distribution
in aiding robustness, which we envisage as working in at least two ways.

The t distribution
has been recommended by Lange {\em et al} (1989), Butler {\em et al} (1990) and Taylor and Verbyla (2004) for robust
least-squares fitting.  As we shall show, our new distribution distorts the results in regression analyses less often than
does the use of the t distribution when the true distribution is Gaussian.
Hence we have protection against outliers, with less chance of distortion
of the results when there are no outliers. 

A second way in which use of this distribution can promote robustness
of results is through sensitivity analysis. When modelling data with the
t distribution, we repeat the analysis using the twin-t distribution.
We have found from looking at a variety of datasets that usually the fit
is only slightly better or worse than the t distribution fit, but the estimated tail behaviour can be different. This type
of sensitivity analysis could be formalised using Bayesian model averaging,
for example, where our uncertainty about the distributional shape can be allowed for.

We call our new distribution the `twin-t' distribution for slightly whimsical reasons that are given in Baker (2014).
We introduce the new distribution, and its statistical properties, in section \ref{sec:2} and then give
examples of its use in regression modelling in section \ref{sec:3}. Azzalini and Genton (2008) have created skewed t distributions as an aid to robust modelling and in section \ref{sec:4} we derive some skew versions of our new distribution. We  apply these skew distributions in section \ref{sec:5}
and in appendices A and B we describe  multivariate and other generalisations. We conclude with a discussion in section \ref{sec:8}.

\section{The twin-t distribution and its properties\label{sec:2}}
\subsection{Motivation}
As is well-known,the t-distribution can be derived by allowing the inverse variance of the normal distribution to be a random variable from a gamma distribution.
We generalise the normal distribution in a quite different way.
The exponential function lies at the heart of definition of the probability density function (pdf) of the normal distribution, and so we derive a new distribution
by generalising the exponential function.
To do this we create
a generalised natural logarithm function
\[\ln_\alpha(x)=\sinh\{\alpha\ln(x)\}/\alpha=\frac{x^\alpha-x^{-\alpha}}{2\alpha}\]
and its inverse, our generalised exponential function
\[\exp_{\alpha}(x)=\exp\{\sinh^{-1}(\alpha x)/\alpha\}=\{\alpha x+\sqrt{1+(\alpha x)^2}\}^{1/\alpha}.\]
These two functions have the correct range, and reduce to the natural logarithm and the
exponential, respectively, as $\alpha\rightarrow 0$.

\subsection{The definition of the kernel of the twin-t distribution}
When generalising
the Gaussian distribution, it is then natural to replace the exponential function
by the generalised exponential function in the definition of the pdf. Hence we can take a  pdf as $f(x) \propto \exp_\alpha(-x^2/2)$,
which motivates the twin-t distribution.
Replacing $\exp(-x^2/2)$ by $\exp_\alpha(-x^2/2)$ we would have the pdf
\[f(x) \propto (-\alpha  x^2/2+\sqrt{1+(\alpha x^2/2)^2})^{1/\alpha} \propto (\alpha  x^2/2+\sqrt{1+(\alpha x^2/2)^2})^{-1/\alpha}.\]
It is convenient to set $\alpha=2/(\nu+1)$ and to rescale the random variable so that $x \rightarrow \sqrt{(\nu+1)/\nu} x$.
This gives the pdf of the twin-t distribution as
\begin{equation}
\label{pdf}
f(x) =k(x^2/\nu+\sqrt{1+(x^2/\nu)^2})^{-(\nu+1)/2}\end{equation}
This definition has the advantage that the distribution
is defined for $\nu > 0$, as is the t distribution, and the twin-t and conventional t distributions have the same power behaviour in the tails. Furthermore  by replacing the term $(x^2/\nu)^2$ in (\ref{pdf}) with zero we obtain the kernel of the conventional t distribution. The constant of normalisation $k$, to be calculated below, ensures that the pdf integrates to one.

The result of using the generalised exponential is not the same
as simply applying Johnson's arcsinh transformation  (\eg Johnson {\em et al}, 1994, Chapter 1) to the normal random
variable or its square, as only the pdf is changed, and there is no Jacobian for the
transformation. Our proposal for the twin-t distribution also differs from the sinh-arcsinh distributions of Jones and Pewsey (2009) for the same reason.

\subsection{Main results and properties}
It is convenient for brevity to define $S=S(x)=x^2/\nu$ and $C=C(x)=\sqrt{1+S^2}$. The
identity $(C+S)=(C-S)^{-1}$ will often be useful.

\subsubsection{The constant of normalisation\label{normcon}}
An essential task is to find the constant of normalisation $k$ for the pdf. The simplest
method is probably to set $(C+S)^{-2}=q$, where $ 0 \le q
\le 1$. Then $C-S=q^{1/2}$, $C+S=q^{-1/2}$, giving $S=x^2/\nu=(1-q)/2q^{1/2}$, from
which for $x > 0$
\[\dee x/\dee q=(-1/4)\sqrt{\nu/2}\frac{1+q}{q^{5/4}(1-q)^{1/2}},\]
allowing the necessary integral to be obtained in terms of the summation of two beta functions, where one evaluates the integral of $f(x)$ over $x > 0$ and then, because the pdf is an even function, doubles the value of this integral to integrate over the entire real line. The value of $k$ can then be obtained as the constant (which depends on $\nu$) necessary to ensure that the pdf integrates to one.

Upon substituting the resulting value of $k$, the pdf is then
\begin{equation}f(x)=\frac{2^{5/2}\Gamma(\nu/4+3/2)}{\sqrt{\pi \nu}\Gamma(\nu/4)(\nu+1)}(x^2/\nu+\sqrt{1+(x^2/\nu)^2})^{-(\nu+1)/2}.\label{eq:t}\end{equation}
As $\nu\rightarrow\infty$ the distribution becomes standard normal.

{\bf Inline supplementary code segment 1 for computing the pdf}
\subsubsection{The cumulative distribution function\label{sec:dist}}
The cumulative distribution function (cdf) for $x > 0$ can be found by writing
\[F(x)-1/2=k\int_0^xf(u)\dee u=k\int_p^1 q^{(\nu+1)/4}|\dee x/\dee q|\dee q,\]
where $p=p(x)=(C(x)+S(x))^{-2}$.
Then
\begin{equation}F(x)-1/2=(k/4)\sqrt{\nu/2}\int_{p(x)}^1(1+q)q^{\nu/4-1}(1-q)^{-1/2}\dee q,\label{eq:f2}\end{equation}
We write $(1+q)q^{\nu/4-1}(1-q)^{-1/2}=q^{\nu/4-1}(1-q)^{1/2}+2q^{\nu/4}(1-q)^{-1/2}$ in the integral immediately above and integrate the second term by parts
(differentiating $q^{\nu/4}$ and integrating $(1-q)^{-1/2}$). On expressing the result in terms of $x$ rather than $p(x)$, using the result $S=x^2/\nu=p^{-1/2}(1-p)/2$, we have that
\begin{equation}F(x)=1+\frac{2^{3/2}x(C(x)+S(x))^{-(\nu+1)/2}}{\sqrt{\nu}(\nu+1)B(\nu/4,3/2)}-\frac{1}{2}I((C(x)+S(x))^{-2};\nu/4,3/2)\label{eq:bigf}\end{equation}
for $x>0$, where $B$ is the beta function and $I$ the (regularised) incomplete beta
function (the distribution function of the beta distribution, $I(z;\alpha,\beta)=\int_0^z q^{\alpha-1}(1-q)^{\beta-1}\dee q/B(\alpha,\beta)$). For $x < 0$, by symmetry, $F(-x)=1-F(x)$. This form is simplest for computation as it only uses one incomplete beta function. The cdf can also be written as
\begin{equation}F(x)=\frac{1}{2}+\frac{2^{3/2}x(C(x)+S(x))^{-(\nu+1)/2}}{\sqrt{\nu}(\nu+1)B(\nu/4,3/2)}+\frac{1}{2}I(1-(C(x)+S(x))^{-2};3/2,\nu/4)\label{eq:bigf2}\end{equation}
using $I(z;\alpha,\beta)=1-I(1-z;\beta,\alpha)$.

{\bf Inline supplementary code segment 2 for computing the distribution function}

{\bf Inline supplementary code segment 3 for computing the quantiles}

{\bf Inline supplementary code segment 4 called from segment 3}
\subsubsection{Moments}
The odd moments about the mean are zero, and the even moments are calculable analytically, using the same type of approach used in section \ref{normcon} to evaluate $k$, but are less friendly than those
for the t distribution.

Thus the variance of the twin-t distribution is
(\ref{eq:t}) is
\begin{equation}\sigma^2=\frac{4(\nu+2)}{(\nu+4)(\nu-2)}(\frac{\Gamma(\nu/4+1/2)}{\Gamma(\nu/4)})^2\label{eq:var}\end{equation}
for $\nu > 2$.

The fourth moment is
\[\text{E}(X^4)=\frac{3\nu^2}{(\nu-4)(\nu+6)}\]
for $\nu > 4$.
In general,
\[\text{E}(X^{2m})=\frac{2^{3-3m}\nu^m (2m-1)!\Gamma(\nu/4-m/2)\Gamma(\nu/4+3/2)}{(\nu+2m+2)(m-1)!\Gamma(\nu/4)\Gamma(\nu/4+m/2+1/2)}\]
for $\nu > 2m$
Also,
\begin{equation}\label{eq:mod}\text{E}(|X|)=\frac{2^{7/2}\nu^{1/2}\Gamma(\nu/4+3/2)}{\sqrt{\pi}\Gamma(\nu/4)(\nu-1)(\nu+3)}\end{equation}
for $\nu > 1$.

\subsection{Comparing the twin-t distribution to the standard normal and t distributions}
\begin{figure}[b]
\centering
\makebox{\includegraphics{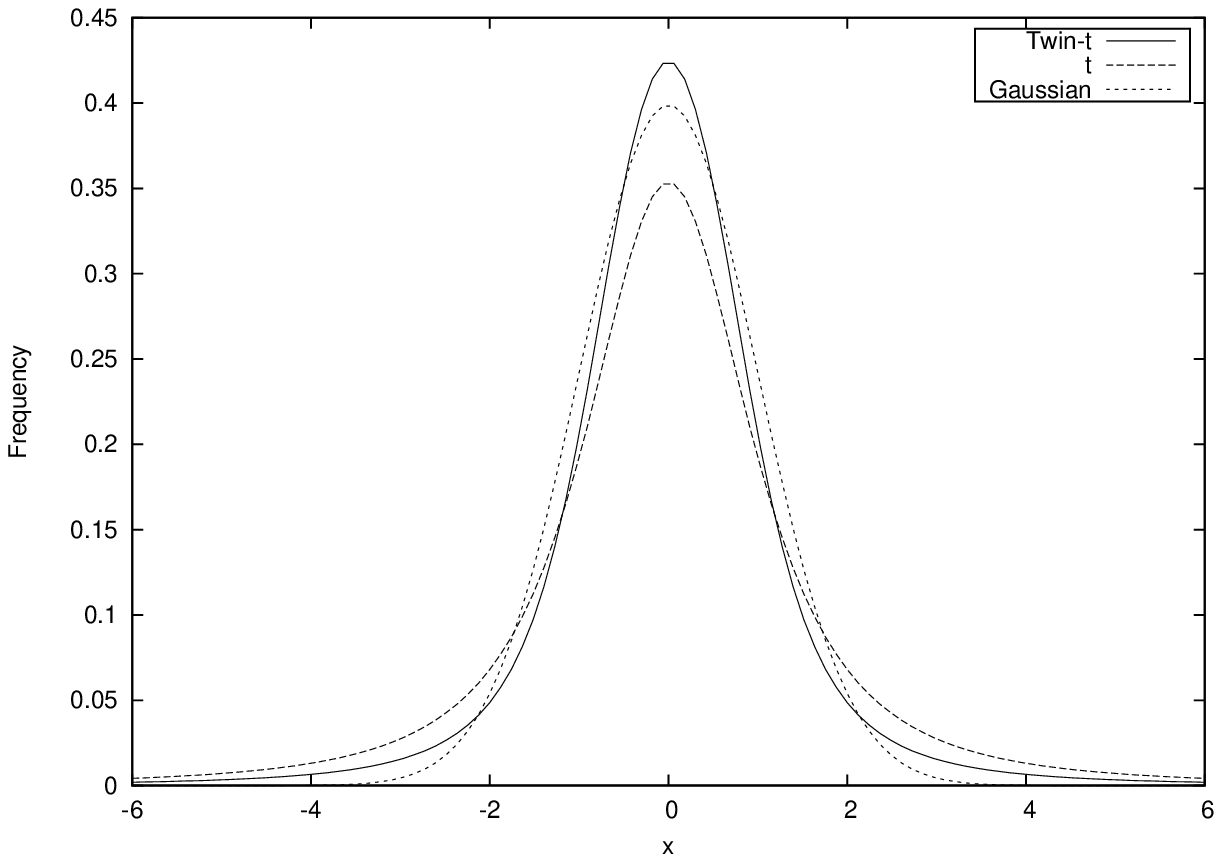}}
\caption{\label{fig:t}The pdf of the twin-t distribution, with $\nu=2$, along
with the t distribution, again with $\nu=2$, and the standard normal distribution.}
\end{figure}

Figure \ref{fig:t} shows a plot of the twin-t distribution, with $\nu=2$, along with the standard normal
and t (with two degrees of freedom) distributions. It can be seen that the distribution is normal-like
in its main body, but with the same power-law behaviour as the t distribution
in the extreme tails.

The reason for this behaviour can be seen from the Taylor series expansion
\begin{equation}
\label{TS}
\ln f(x) =\ln(k)-\frac{\nu+1}{2}(x^2/\nu-(1/6)(x^2/\nu)^3+(3/40)(x^2/\nu)^5\cdots).\end{equation}
The first correction to normality is of order $x^6$, not of order $x^4$
as it is for the t distribution. The expansion (\ref{TS}) can most easily be obtained by writing the density of the
twin-t distribution in terms of the arcsinh function and using this function's Taylor Series expansion.

Also from (\ref{TS}), in the main body of the distribution the pdf
has the shape of a normal distribution, with variance $\nu/(\nu+1)$. This suggests that the distribution could be useful when one essentially
wants a normal distribution, but with robustness to outliers. Applications
could be robust least-squares fitting (cf Lange {\em et al}, 1989), prior
distributions, or for importance sampling. Here it is often required
to produce a weighted sample from a normal distribution, but to be efficient
the sampling distribution must be fatter in the tails than the target
distribution.

The t distribution can be generated by making the inverse variance $w=1/\sigma^2$ of the normal distribution
a gamma-distributed random variate. We naturally ask what pdf $g(w)$ would convert the normal distribution
into the twin-t distribution, so that 
\begin{equation}\frac{1}{2\pi}\int_0^\infty w^{1/2}\exp(-x^2 w/2)g(w)\dee w=k(x^2/\nu+\sqrt{1+(x^2/\nu)^2})^{-(\nu+1)/2}.\label{eq:lap}\end{equation}
Writing $x^2/2=s$, the twin-t pdf expressed in terms of $s$ is the Laplace transform of $w^{1/2}g(w)$, from which using the table of Laplace transforms in Schiff
(1999),p 214, we have $g(w) \propto J_{(\nu+1)/2}(2w/\nu)/w^{3/2}$, where $J$ is the Bessel function of the first kind.
Bessel functions oscillate in sign, so $g(w)$ is not a pdf, and hence (\ref{eq:lap}) does not admit a probabilistic interpretation.
The twin-t distribution cannot be derived from the Gaussian distribution by making the variance a random variable. Also, it is not possible to generate random variates from the twin-t
distribution by generating Gaussian random variables and randomly rescaling them.
 
\subsection{Random numbers}
One possible way to generate random numbers from this distribution would
be by rejection sampling from a t distribution. Denoting the pdfs of
the new distribution and the t distribution with $\nu$ degrees of freedom
by $f(x)$, $g(x)$ respectively, one takes a random variable $X$ from
the t distribution, and accepts it with probability
\begin{equation}\frac{f(X)}{g(X)}/\frac{f(0)}{g(0)},\label{eq:rand}\end{equation}
\ie with probability
\[p=\{\frac{1+X^2/\nu}{\sqrt{1+(X^2/\nu)^2}+X^2/\nu}\}^{(\nu+1)/2}.\]

The overall probability of acceptance is $\int_{-\infty}^\infty\frac{f(x)}{g(x)}/\frac{f(0)}{g(0)} g(x)\dee x= g(0)/f(0)$, so that 
on average 1.271 random
numbers from the t distribution are used to generate a random number from the twin-t distribution when $\nu=1$, declining to 1.035
when $\nu=20$. Random numbers from the t distribution can be generated \eg using the polar method of Bailey (1994).

{\bf Inline supplementary code segment 5 for computing random numbers}
\subsection{Miscellaneous properties}
The distribution function takes a relatively simple form when $\nu=2$
and still more so when $\nu=4$.

When $\nu=2$, from (\ref{eq:bigf}) or directly from (\ref{eq:f2}),
\[F(x)=1+ (1/3\pi)x(C+S)^{-3/2}-(1/\pi)\sin^{-1}(1/(C+S)),\]
where $C=\sqrt{1+(x^2/2)^2}$, $S=x^2/2$.

When $\nu=4$,
\[F(x)=1/2+(3/5)2^{-1/2}x(\sqrt{1+(x^2/4)^2}+x^2/4)^{-5/2}+2^{-5/2}x^3(\sqrt{1+(x^2/4)^2}+x^2/4)^{-3/2}.\]
From (\ref{eq:var}), the variance is $3\pi/8$. In general, when $\nu=4m$,
the pdf takes a simple form and the distribution function can be calculated
in terms of elementary functions.

\subsection{Negative degrees of freedom}

Papastathopoulos and Tawn (2013) showed that when $\nu < 0$ the conventional definition of the t distribution with positive degrees of freedom can be extended to give rise to a short-tailed
distribution, where $x < |\nu|$. Here however this is not the
case; because $C+S=(C-S)^{-1}$, for $\nu=-\xi$ we obtain the same distribution
as for $\nu=\xi-2$, but where the random variable has been rescaled. When
$\xi \le 2$, $\nu \le 0$ and the integral of the pdf does not converge.
Hence here the case $\nu < 0$ gives nothing new.

\section{Application: robust regression\label{sec:3}}
Perhaps the most obvious application of the twin-t distribution is its use in robust regression. Standard
linear regression assumes that the errors are normally distributed but
the t distribution is an alternative  that can be used for
this purpose in order to model and downweight outlying or unusual observations
(Taylor and Verbyla, 2004). When using the t distribution for modelling the errors, instead of assuming these are independently and identically distributed (IID) as $\epsilon_i \sim \sigma N(0,1)$ as in standard linear regression, we assume these are IID   $\epsilon_i \sim \sigma t_\nu$, where $t_\nu$ denotes a t distribution with $\nu$ degrees of freedom. Similarly when using the twin-t distribution to model the errors, we assume that these are IID $\epsilon_i \sim \sigma tt_\nu$, where $tt_\nu$ denotes a twin-t distribution. The pdf of the errors are then given by $\sigma^{-1}f(x/\sigma)$, where $f(\cdot)$ is given in (\ref{eq:t}), so that the regression's likelihood can easily be calculated and the standard asymptotic theory of maximum likelihood can be used to make inferences.

\subsection{A simulation study\label{simstudy}}
Following the simulation study design of Taylor and Verbyla (2004) we simulated
bivariate data where, in a dataset of size $n$, the covariates were calculated
deterministically as $0.1 + 9.9(i-1)/(n-1)$, for $i=1, \cdots, n$, \ie evenly spaced from 0.1 to 10. The
outcomes were then simulated as
\[
Y_i|X_i = \beta_0+\beta_1 X_i + \epsilon_i
\]
Continuing to follow the simulation study from Taylor and Verbyla,  for
all simulated datasets we used  $\beta_0=-0.5$ and $\beta_1=2$. Also as
in Taylor and Verbyla, we simulated datasets where the error terms follow
a t distribution with either 3, 5 or 8 degrees of freedom. However we
also simulated datasets where the error terms follow a standard normal
distribution, ie we also simulated data from a t distribution with infinite
degrees of freedom. Like Taylor and Verbyla, we used $n=50, 100, 200$,
providing three different sample sizes and so, with four alternative degrees
of freedom, 12 scenarios were explored. We simulated 1000 datasets for
each scenario. The simulation study was performed using $R$ software.

We fitted three models to each simulated dataset: 1) a standard linear
regression using the `lm' command 2) a linear model where the error term is
assumed to follow a t distribution (using the {\it hett R} package; a `t distribution regression')
and 3) a linear model where the error term is assumed to follow a the twin-t distribution
distribution (a `twin-t distribution regression'). The last of these three models was fitted by computing the
likelihood in terms of the regression parameters $\beta_0$ and $\beta_1$, $\nu$ and $\sigma$, and maximising this likelihood using $R$'s in-built optimiser {\it optim}.
For models 2 and 3, both $\sigma$ and $\nu$ were
estimated by maximum likelihood along with the two regression parameters, where the likelihood was parameterised  on the $\log(\sigma)$ and $\log(\nu)$ scales so that no constraints on the parameter space were required when performing the maximisation.

Our intuition is that the linear model using the twin-t distribution
for the error terms will perform more like a standard linear model (than
a `t distribution regression') when the normality assumption
is true, but also that the twin-t distribution will behave similarly
to a t distribution when this is instead the true model.

\subsubsection{Results when the normality assumption is true}
\begin{figure}
\centering
\includegraphics[width=0.7\textwidth]{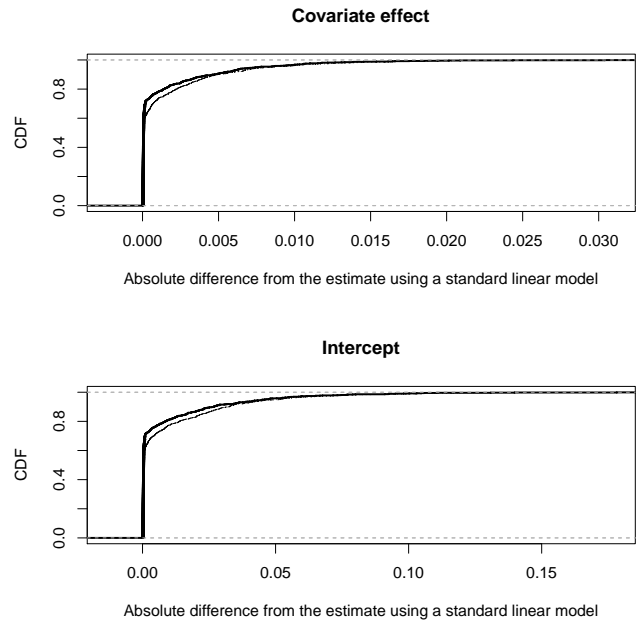}
\caption{\label{figreg} The results for $n=100$ when the
normality assumption is true, where the top panel shows the
results for the estimated covariate effect and the bottom panel
shows the results for the intercept.  The thicker lines show the
cdf of the absolute differences between the twin-t distribution regression
estimates and the standard linear regression estimates, and the
thinner line shows the cdf of the absolute differences between
the t distribution regression estimates and standard linear
regression estimates.}
\end{figure}

For each imputed dataset in scenarios where the normality assumption
is true, we calculated the absolute difference between the twin-t distribution
regression estimates (both the covariate effect and the intercept) and
the standard linear regression estimates. We then plotted the cdf of these absolute differences in each of these
scenarios. For comparison we repeated this exercise using the absolute
difference between the t distribution regression estimates and the standard
linear regression estimates. The results are shown for $n=100$ in figure
\ref{figreg}, where the thicker lines show the cdf of the absolute differences
between the twin-t regression estimates and the standard linear regression
estimates, and the thinner line shows the cdf of the absolute differences
between the t distribution regression estimates and standard linear regression
estimates.

From figure \ref{figreg} we see that in around 60\% of simulated datasets
the absolute difference between the standard linear regression estimates and
both the twin-t and the t distribution regression estimates are effectively zero;
in quite a large proportion of datasets both the twin-t and the t distribution regressions
replicate the standard linear regression estimates. However the twin-t distribution regression
replicates the standard linear regression estimates more often, and around 75\% of
the time (figure \ref{figreg}). Similar observations were also made for $n=50, 200$ (results not shown).
\subsubsection{Results when the normality assumption is false}
\begin{figure}
\begin{center}
\includegraphics[width=0.7\textwidth]{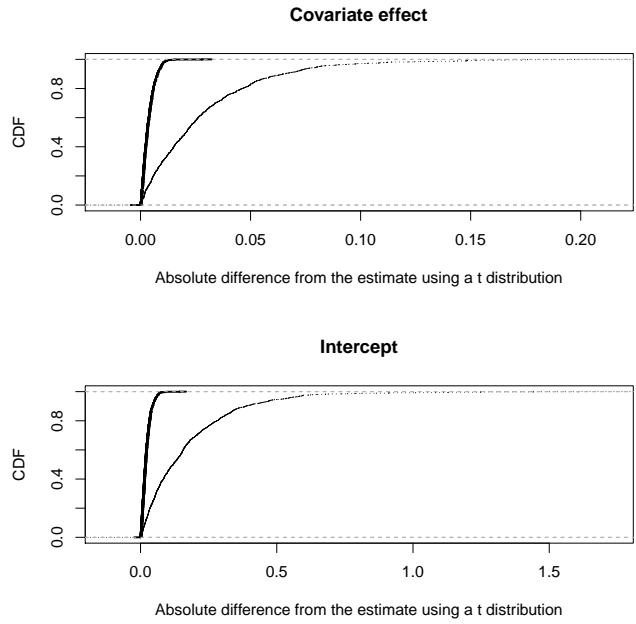}
\caption{\label{fig2} The results for $n=100$ when the t distribution with 3 degrees of freedom is true,
where the top panel shows the results for the estimated covariate effect and the bottom panel shows the results for
the intercept. The thicker lines show the cdf of the absolute
differences between the twin-t distribution regression estimates
and the t distribution regression estimates, and the thinner line shows the cdf of the absolute differences between
the standard linear regression estimates and t distribution regression estimates.}
\end{center}
\end{figure}

Another intuition is that twin-t regressions will behave similarly
to t distribution regressions when the latter's assumptions  are true.  Hence for the scenarios where
the t distribution is true we calculated the absolute difference between
the twin-t distribution  regression estimates and the  t distribution regression estimates. For comparison, we also calculated
the absolute differences between the standard linear regression estimates
and the t distribution regression estimates. Again we plotted the cdfs
of these absolute differences and the results for $n=100$ and 3 degrees
of freedom are shown in figure \ref{fig2}.

Figure \ref{fig2} shows that the twin-t regression behaves similarly
to a t distribution; the estimates produced do not differ greatly from
those from a t distribution, particularly when we compare its performance
to a standard linear regression. Similar observations were also made for $n=50, 200$ (results not shown)

To summarise, the simulation study supports our intuition: the twin-t distribution
regression usually behaves like a standard linear regression  when the
normality assumption is true and also behaves similarly to a t distribution regression
when this is the correct model.

\subsection{Example: Martin Marietta data}
This example was taken from Taylor and Verbyla (2004) who also use this example in their section 5.1. Here the  excess rate of returns $Y$ of the Marietta company,
(a consumer product manufacturing, development, and design company) is regressed on excess rate of return for the New York Exchange ($X$), where the rates of return were measured
monthly, from 1982 to 1986. The data were obtained from the R package {\it hett} which uses this example as an example dataset. This example's residuals have heavy tails which motivates the use of alternatives to standard linear regressions. 

Taylor and Verbyla (2004) suggest an extension of the type of modelling used in the simulation study when modelling these data, where heteroscedasticity is allowed by assuming that $\log(\sigma^2)=\lambda_0+\lambda_1 x_i$, so that the amount dispersion depends on the covariate. By constraining $\lambda_1=0$ we obtain the model used previously in section \ref{simstudy}. The likelihood function used in the simulation study was modified slightly to accommodate this extension and inferences using maximum likelihood were obtained using the {\it maxLik} package.

Very similar inferences for the regression parameters were obtained to those shown using a t distribution for the errors in table 4 of Taylor and Verbyla (2004). Hence this example provides a nice example where using the twin-t distribution in a sensitivity analysis supports the conclusions from an established alternative. However the twin-t distribution appears to provide a slightly better fit for these data. For the homoscedastic regression models ($\lambda_1=0$) using the {\it hett} package to fit a t distribution regression achieved a maximum log likelihood of $\ell = 71.81$ and the corresponding twin-t regression achieved $\ell = 72.40$. For the heteroscedastic regression models ($\lambda_1 \ne 0$) the  t distribution regression achieved a maximum log likelihood of $\ell = 73.37$ but the corresponding twin-t regression achieved $\ell = 73.48$. AIC statistics therefore support the use of the twin-t distribution to model these data but the improvement in model fit is modest.

\section{Skewing the distribution\label{sec:4}}
We discuss this before describing curve fitting, as the majority of datasets do have
at least slightly skew distributions. There are many ways of skewing distributions and we explored three possibilities.
The simplest is the 2-piece skew distribution, where the scale of the random variable differs between the two halves. 
A more complex method gave results similar to the Jones and Faddy (2003) skew t distribution. Finally an Azzalini-type skew distribution was developed.
Although the two latter approaches were much more complex than using the 2-piece idea, it was found after fitting a number of datasets that the 2-piece distribution always gave the
best fit, despite its much greater simplicity, and the unsatisfactoriness of a discontinuity in a second derivative at zero. Hence we only briefly touch on our work on the other skew distributions.

Inferences concerning central tendency are perhaps most easily described for the mode of the distribution, which is now different from the mean. Only the standardised distributions are described in the text,
but for data fitting we must introduce both a scale parameter $\sigma$, as used in the regression modelling in section \ref{sec:3}, and a location parameter  $\mu$, so for an observation $Y$ the standardised variable is $X=(Y-\mu)/\sigma$.

If the skew distributions were used for regression modelling, if the mode is $M({\bf z})$, where ${\bf z}$ is a vector of covariates, then the standardised mode is $x_m({\bf z})=(M({\bf z})-\mu)/\sigma$.
The mode is therefore computed as a function of ${\bf z}$ and the regression coefficients. Then from the standardised mode, for the Jones-Faddy and Azzalini style distributions, the skewness parameter, $a, b$ or $\phi$ respectively, can be readily
computed and hence the likelihood function can be computed. For the standardised 2-piece distribution, the mode is at zero, so one simply regresses the location parameter $\mu$ (the mode) on ${\bf z}$.
\subsection{A 2-piece distribution}
We focus
first on the method of forming a 2-piece distribution given in
Fern\'{a}ndez and Steel (1998), and first described by Hansen (1994).
The pdf in (\ref{eq:t}) becomes
$g(x)=\frac{2}{\gamma+\gamma^{-1}}f(x/\gamma)$ for $ x > 0$ and $g(x)=\frac{2}{\gamma+\gamma^{-1}}f(\gamma x)$ for $x < 0$.
This skew distribution is continuous at $x=0$ and has continuous
first derivative (zero) but the second derivative is not
continuous. This can be a problem for frequentist inference with 2-piece distributions;
the use of the Hessian in deriving standard errors on model
parameter estimates is problematical, and one would need to use
the bootstrap. Function minimizers might have problems (not in our experience), but
these are usually robust; otherwise Nelder's simplex method
could be used.

This distribution is mathematically simple and seems to fit well to data.
Its only flaw is the discontinuous second derivative. 

The $r$th moment is $E_s(X^r)=M_r\frac{\gamma^{r+1}-(-1/\gamma)^{r+1}}{\gamma+1/\gamma}$,
where $M_r=\int_0^\infty f(s)\dee s$.
Thus the mean is $\text{E}_s(X)=(\gamma-1/\gamma)\text{E}(|X|)$,
where $\text{E}(|X|)$ is given in (\ref{eq:mod}).

Random numbers can be generated by a simple modification of the rejection method used for the symmetric distribution.
The probability that $X > 0$ is $\gamma^2/(1+\gamma^2)$, so the steps are:
\begin{enumerate}
\item generate a random variable from the t distribution with $\nu$ degrees of freedom and ignore the sign;
\item choose sign of variable positive with probability $\gamma^2/(1+\gamma^2)$;
\item accept r.v. $X$ with probability $p$ as before, \ie
\[p=\{\frac{1+X^2/\nu}{\sqrt{1+(X^2/\nu)^2}+X^2/\nu}\}^{(\nu+1)/2};\]
\item scale variable to $Y=\gamma X$ if $X > 0$, else to $Y=X/\gamma$;
\end{enumerate}

\subsection{A Jones and Faddy (2003) style skew distribution}
Jones {\em et al} (2003) used an ingenious method to skew the t distribution, which can be adapted for
use here.

This approach turned out to be the most complex, and the resulting distribution
has the drawback that the normalising factor for the pdf must be found by numerical integration.
However, the mode is readily computable, and the other model parameters readily found
given the predicted mode, so it can be used for regression modelling as well as curve fitting.

We start from the original pdf $f(x)=c(C+S)^{-(\nu+1)/2}$.
Write
\[p^{1/4}=2^{-1/4}\frac{(C^{1/2}+x/\sqrt{\nu})^{1/2}}{(C+x^2/\nu)^{1/4}},\]
\[r^{1/4}=2^{-1/4}\frac{(C^{1/2}-x/\sqrt{\nu})^{1/2}}{(C+x^2/\nu)^{1/4}}.\]
Then $(pr)^{1/4}=2^{-1/2}/(C+S)$,
\begin{equation}p=1/2+\frac{C^{1/2}x/\sqrt{\nu}}{C+x^2/\nu},\label{eq:px}\end{equation}
and $p+r=1$, so that $r=1-p$.
Hence $p$ and $1-p$ are positive factors where $0 < p < 1$, and $p(-x)=1-p(x)$.
Now $p(x)$ and $(1-p(x))$ can be raised to different powers to obtain a skew distribution.

Hence we define the skew pdf
\begin{equation}f_s(x)=c 2^{(\nu+1)/4}p(x)^{(a+1/2)/4}(1-p(x))^{(b+1/2)/4},\label{eq:fs}\end{equation}
where $a+b=\nu$ and $c$, the constant of normalisation, has yet to be determined. The constant $c$ could be evaluated using (\ref{eq:fs}) directly and numerical integration.

\subsection{An Azzalini-type skew distribution}
Azzalini and Capitiano (1999) and (2003)  describe skew distributions with pdf $2G(\alpha x)f(x)$, where $f$ is a symmetric pdf and $G$ the distribution function of a symmetric distribution.
The skew-t distribution was obtained from the skew-normal distribution, on giving the variance a gamma distribution.
Here we do not have such a genesis of the twin-t distribution, so we use a mathematically convenient distribution function,
derived from the probability $p(x)$ given by (\ref{eq:px}), in fact
\begin{equation}G(x;\phi)=(1-\phi)/2+\phi p(x),\label{eq:azz}\end{equation}
where $-1 < \phi < 1$. Then the skew distribution has pdf $f_s(x)=2G(x;\phi)f(x)$. This has the advantage that the pdf is little more complicated than $f(x)$ and requires no
further computation of special functions or integrals. The even moments are unchanged, and the odd moments are calculable.

The only drawback is that the amount of skewness that can be accommodated is limited; however, this did not pose a problem for the datasets we examined.
In fact, as $p \rightarrow 1$, $1-p \sim x^{-8}$, so this distribution can accommodate $f \sim x^{-\nu}$ in one tail and $f \sim x^{-\nu-8}$ in the other
when $\phi=\pm 1$, which allows sufficient asymmetry.

\section{Curve fitting examples\label{sec:5}}
We now fit all three types of skew twin-t distributions to some contrasting datasets. Here there are no covariates and we describe model frequency distributions of the datasets. In each case we estimate $\mu$, $\sigma$, $\nu$, and the parameters that describe the extent of skewing, using maximum likelihood.
\subsection{Glass fibre strength}
\begin{figure}
\centering
\makebox{\includegraphics{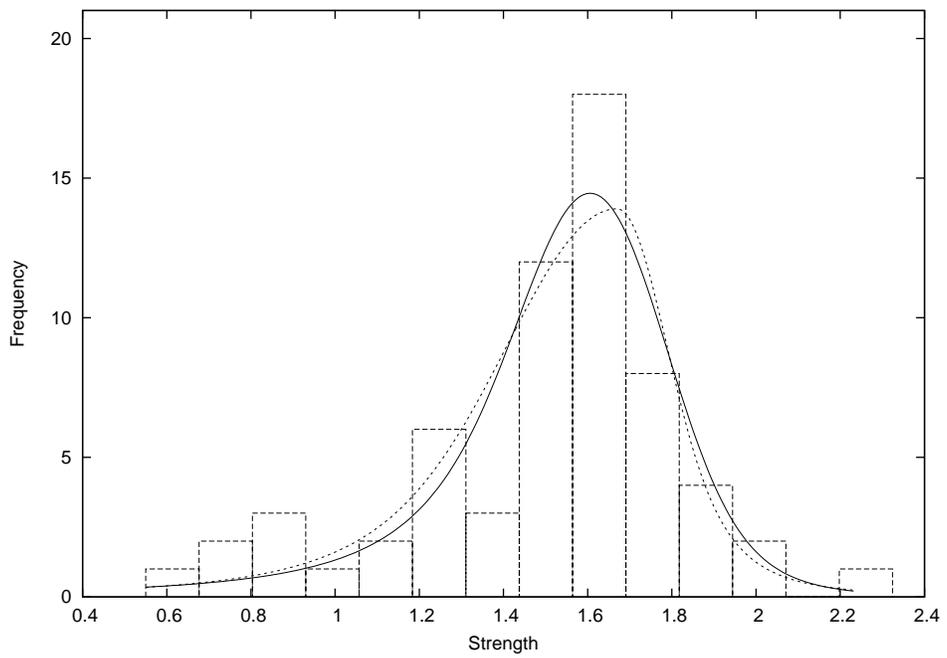}}
\caption{\label{fig:glass}The Jones-type skewed twin-t distribution fitted to the 63 glass fibre breaking strengths in
the Smith and Naylor (1987) dataset. The dotted line is the 2-piece twin-t distribution.}
\end{figure}

Jones and Faddy (2003) showed a fit to the Smith and Naylor (1987) glass fibre strength dataset. This is a nice dataset because it is skew to the left; positive random variables are often skew to the right,
when there are many survival distributions that can be fitted, but these cannot accommodate skewness to the left.
Figure \ref{fig:glass} shows fits of the Jones-type skewed distribution and the 2-piece distribution
to these data. The log-likelihoods for the three skewed distributions in the text were: 2-piece $-\ell=12.599\tilde (\hat{\gamma}=0.67)$, Jones-style distribution $-\ell=13.030$,
Azzalini-style $-\ell=12.622\tilde (\hat{\phi}=-0.878)$.

\subsection{Nitrogen dioxide concentrations}
\begin{figure}
\centering
\makebox{\includegraphics{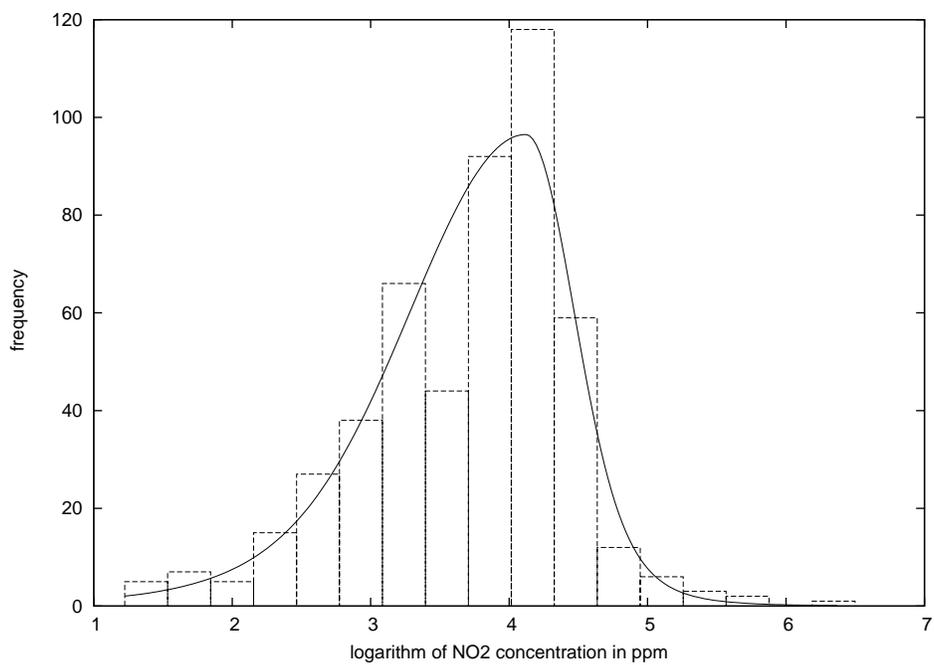}}
\caption{\label{fig:n02}The 2-piece twin-t distribution fitted to the 500 logged $\text{NO}_2$
concentrations in
the  Magne Aldrin dataset, submitted to Statlib in 2004.}
\end{figure}

Another left-skewed dataset is a study submitted to Statlib by
Magne Aldrin (2004), and available from there. Here 500 logged
hourly nitrogen dioxide concentrations are recorded in Oslo in
2001--2003. There are several predictor variables such as
traffic density, but here we simply plotted the distribution and
the fit of the 2-piece distribution. The twin-t distribution
fitted slightly better than the t distribution ($-\ell=560.90$
as opposed to $-\ell=561.21$) but only the 2-piece skewed
distribution fitted the data well ($-\ell=540.08$ as opposed to
$-\ell=553.12$ for the Jones-style skewed distribution). Figure
\ref{fig:n02} shows the fitted curve.

\subsection{FTSE-100 logged daily returns}
The FTSE-100 logged daily returns from April 1984 to July 2013 were also fitted. Here the t distribution gave $-\ell=10668.23$, and the twin-t distribution gave $-\ell=10668.64$, so
the fits are almost identically good. On skewing the twin-t distribution, the 2-piece distribution gave $-\ell=10659.92$, the Jones-style distribution gave $-\ell=10662.22$, and
the Azzalini-style distribution gave $-\ell=10660.81$. Clearly the skewness parameter is significant, the returns in fact being skew to the left, \ie one is more likely to
see a large drop in the FTSE than a large rise.

To show the usefulness of this fitting for sensitivity analysis, the parameter $\nu$ that determines tail behaviour
had 95\% confidence interval $3.54-4.24$ on fitting the t distribution, but $2.65-3.07$ on fitting the twin-t distribution. No statistician would
seek to study tail behaviour by fitting the whole distribution like this, but this example shows how fresh light can be thrown on the accuracy of a
parameter estimate by fitting the twin-t distribution. The 95\% confidence intervals do not even overlap, and fitting the twin-t distribution would reveal to the
practitioner that they did not in reality have the accurate estimate of tail behaviour that they might have previously thought.

There is a corresponding multivariate distribution, given in appendix A. McDonald and Newey (1988) generalised the kernel of the t-distribution to 
$(1+|x|^{\beta}/\gamma)^{-\gamma-(1/{\beta})}$. A corresponding generalisation of the twin-t distribution is given in appendix B.

\section{Conclusions\label{sec:8}}
A distribution has been introduced that has similar tail behaviour to the t distribution, but which is more Gaussian in the centre of the distribution.
It is envisaged that this distribution will be useful in data modelling, piggybacking on the widespread use of the t distribution itself.
Its usefulness has been illustrated by examining its performance in robust regression in which, as expected, outliers can be coped with if present; and if absent, the results
are more often similar to an ordinary least squares fit than they would be if a t distribution were used.

Continuing the theme of robustness, the distribution could also be used instead of a t distribution as part of a sensitivity analysis, or via Bayesian model averaging
along with the t distribution, if we wish to capture our uncertainty about the shape of the distribution. The example of finding the tail behaviour of the distribution of FTSE returns
showed how use of this distribution can reveal some of the true uncertainty about the value of a model parameter.

Some effort has been devoted to skewing the distribution,
as distributions of data are often at least slightly skew, and robust inference must allow some skewness. Here we have devised three skew distributions, the 2-piece distribution, a distribution derived using the approach of Jones and Faddy (2003), and one derived using the approach of Azzalini
and Capitiano (2003).
For its simplicity, and because it always fitted well in our
experience, we recommend the simple and reliable 2-piece twin-t distribution.

We hope that there will be many uses for this new distribution and its skewed variant. Further work could focus on applications.
To facilitate this, an R package that provides the usual functionality (pdf, distribution function, inverse distribution function and random numbers)
is available from the authors, and supplementary code that does this is provided with the online version of this paper.

\clearpage
\section*{Appendix A:The multivariate case}
There is a corresponding multivariate distribution; we first consider the standard distribution for a p-dimensional random variate
${\bf Y}$ where the $Y_i$ are uncorrelated, with pdf
\[f({\bf y})=k\{{\bf y}^T{\bf y}/\nu+\sqrt{1+({\bf y}^T{\bf y}/\nu)^2}\}^{-(\nu+p)/2}.\]
The power $(\nu+p)/2$ is chosen to make the pdf defined for $\nu > 0$.
The constant of proportionality is found by changing variables of integration
to a radius $r$ where $r^2={\bf y}^T{\bf y}$ and polar coordinates. Then since the surface area of
a $p$-dimensional ball is $p\pi^{p/2}r^{p-1}/\Gamma(p/2+1)$, the integral
gives
\[k^{-1}=\frac{p\pi^{p/2}}{\Gamma(p/2+1)}\int_0^\infty
\frac{r^{p-1}\dee r}{(r^2/\nu+\sqrt{1+(r^2/\nu)^2})^{(\nu+p)/2}}.\]

To obtain the pdf for the general distribution with location and scale parameters, we write ${\bf Y}={\bf H}({\bf X}-{\boldsymbol \mu})$, where we define ${\bf V}^{-1}={\bf H}^T{\bf H}$.
Then $\prod_{i=1}^p \dee y_i=|{\bf H}|\prod_{i=1}^p\dee x_i$, \ie $\dee {\bf y}=|{\bf H}|\dee {\bf x}=|{\bf V}|^{-1/2}\dee {\bf x}$.
One can therefore write the
general $p$-dimensional pdf $f({\bf x})$ as
\[\frac{k}{|{\bf V}|^{1/2}} \{({\bf x}-{\bf \mu})^T{\bf V}^{-1}({\bf x}-{\bf \mu})/\nu+\sqrt{1+(({\bf x}-{\bf \mu})^T{\bf V}^{-1}({\bf x}-{\bf \mu})/\nu)^2}\}^{-(\nu+p)/2}.\]
Using the same method of evaluating the integral for $k$ as in the univariate case, after simplifying the pdf $f({\bf x})$ is
\begin{equation}
\frac{2^{2+p/2}\Gamma(\nu/4+p/2+1)}{|{\bf V}|^{1/2}(\nu \pi)^{p/2}\Gamma(\nu/4)(\nu+p)} \{({\bf x}-{\bf \mu})^T{\bf V}^{-1}({\bf x}-{\bf \mu})/\nu+\sqrt{1+(({\bf x}-{\bf \mu})^T{\bf V}^{-1}({\bf x}-{\bf \mu})/\nu)^2}\}^{-(\nu+p)/2}.\label{eq:mult}\end{equation}
When $p=1$ we regain the univariate pdf in (\ref{eq:t}).

As with the multivariate t distribution, the random variables of the standard distribution are uncorrelated but not independent.
Hence $\text{E}(Y_iY_j)=0$ if $i \ne j$, and $\text{E}(Y_i^2)=\text{E}(r^2)/p$.
Since $\text{E}(r^2)$ can readily be found, so can the second moments of the standard distribution. On again using ${\bf Y}={\bf H}{\bf X}$
we obtain the moments of the distribution of ${\bf X}$ as 
\[\text{E}\{(x_i-\mu_i)(x_j-\mu_j)\}=\text{E}\{(r^2)\}V_{ij}/p|{\bf V}|^{1/2}.\]
On evaluating $\text{E}(r^2)$ one obtains for the second central moment
\[\text{E}\{(x_i-\mu_i)(x_j-\mu_j)\}=\frac{\nu}{\nu-2}\frac{\Gamma(\nu/4+1/2)\Gamma(\nu/4+p/2+1)}{\Gamma(\nu/4)\Gamma(\nu/4+p/2+3/2)}V_{ij}.\]
This gives a simple result for the bivariate case $p=2$, when
\[\text{E}\{(x_i-\mu_i)(x_j-\mu_j)\}=\frac{\nu^2(\nu+4)}{(\nu+6)(\nu+2)(\nu-2)}V_{ij}.\]

These results mean that, when fitting this model to data, good starting
values for a likelihood maximisation can be obtained by fixing $\nu$
and using the method of moments, if $\nu > 2$.
\section*{Appendix B: Generalisations}
McDonald and Newey (1988) introduced the generalised t
distribution, with the kernel of the pdf as
$(1+|x|^{\beta}/\gamma)^{-\gamma-(1/{\beta})}$, where $\beta >
0$, which is a rescaled t distribution when $\beta=2$.
This generalisation has been useful in finance. It is possible to generalise the twin-t distribution similarly, to give a distribution with pdf
\begin{equation}f(x)=\frac{(\sqrt{1+(|x|^{\beta}/\gamma)^2}+|x|^{\beta}/\gamma)^{-\gamma-1/{\beta}}}{(\gamma/2)^{1/{\beta}}(\gamma+1/{\beta})B(\gamma/2,1/{\beta}+1)},\label{eq:newey}\end{equation}
where the normalisation factor is derived in the usual way by setting $q=(\sqrt{1+(|x|^{\beta}/\gamma)^2}+|x|^{\beta}/\gamma))^{-2}$, when $x^{\beta}/\gamma=(q^{-1/2}-q^{1/2})/2$,
from which the Jacobian of the transformation to $q$ is $|\dee x/\dee q|=(1/2{\beta})(\gamma/2)^{1/{\beta}}(1+q)q^{-1/2{\beta}-1}(1-q)^{1/{\beta}-1}$, and on evaluating the sum of the
two beta function integrals and simplifying the result, we obtain (\ref{eq:newey}).
When $\beta=2$ we regain the twin-t distribution, but with rescaled random variable, so that the random variable is now $2^{1/2}$ times that in (\ref{eq:t}).

The distribution function follows on proceeding as in section \ref{sec:dist} and is for $x > 0$
\begin{equation}F(x)=1+\frac{x(\sqrt{1+(|x|^{\beta}/\gamma)^2}+|x|^{\beta}/\gamma)^{-\gamma-1/{\beta}}}{(\gamma/2)^{1/{\beta}}(\gamma+1/{\beta})B(\gamma/2,1/{\beta}+1)}-(1/2)I(\gamma/2,1/{\beta}+1).\label{eq:mdist}\end{equation}
The general (noninteger) moment $\text{E}(|X|^r)$ is
\[\text{E}(|X|^r)=\frac{(\gamma/2)^{r/{\beta}}(\gamma+2/{\beta})B(\gamma/2-r/2{\beta},(r+1)/{\beta})}{(\gamma+r/{\beta}+2/{\beta})B(\gamma/2,1/{\beta})}\]
for $r < {\beta}\gamma$.

For completeness, we mention the corresponding multivariate distribution, with pdf
\begin{eqnarray*}f({\bf x})&=&\frac{\beta\Gamma(p/2+1)(\gamma+2p/\beta)}{(\gamma+p/\beta)(\gamma/2)^{p/\beta}p\pi^{p/2}B(\gamma/2,p/\beta)|{\bf V}|^{1/2}}\\
&&((({\bf x}-{\bf \mu})^T{\bf V}^{-1}({\bf x}-{\bf \mu}))^{\beta/2}/\gamma+\sqrt{1+\{(({\bf x}-{\bf \mu})^T{\bf V}^{-1}({\bf x}-{\bf \mu}))^{\beta/2})/\gamma\}^2})^{-\gamma-p/\beta},\end{eqnarray*}
where the notation is as in appendix A and $p$ is the dimensionality  of ${\bf x}$.

This is derived in the same way as the multivariate distribution in appendix A
\clearpage
\section*{Inline supplementary code}
\begin{verbatim}
dtt <-function (x, df) 
{
constant = 2^(3/2)/((df^0.5) * (df + 1) * beta(df/4, 3/2))
kernel=x^2/df
constant*(exp(asinh(kernel)))^(-(df+1)/2)
}
\end{verbatim}
Inline code segment 1: R code for the pdf of the twin-t distribution.

\begin{verbatim}
ptt <-function (x,  df) 
{
S=x^2/df; C=(1+S^2)^0.5; p=C+S
F=1+(2^(3/2)*abs(x)*p^(-(df+1)/2))/(df^0.5*(df+1)*beta(df/4, 3/2))-
0.5*pbeta(p^(-2), df/4, 3/2)
F[x<0]=1-F[x<0]
F
}
\end{verbatim}
Inline code segment 2: R code for the distribution function of the twin-t distribution.
\begin{verbatim}
qtt <-function (x, df) 
{
y=NULL
if (length(df)==1) df=rep(df, length(x))
for(i in 1:length(x))
y[i]=NR(x[i], df[i])
y
}
\end{verbatim}
Inline code segment 3: R code for the quantiles of the twin-t distribution.
\begin{verbatim}
NR <-function (x, v, tol=1e-10) 
{
diff=Inf; x_old=0
while(diff>tol)
{
x_new=x_old-(ptt(x_old, v)-x)/dtt(x_old,v)
diff=abs(x_new-x_old)
x_old=x_new
}
x_new
}
\end{verbatim}
Inline code segment 4: R code called by the previous R function.
\begin{verbatim}
rtt <-function (n, df) 
{
if (length(df)==1) df=rep(df, n)
y=NULL
while(length(y)<n) {
v=df[length(y)+1]
x=rt(1, df=v)
S=x^2/v
p=((1+S)/((1+S^2)^0.5+S))^((v+1)/2)
r=runif(1)
if(r<p) y=c(y,x)
}
y
}
\end{verbatim}
Inline code segment 5: R code for generating random numbers from the twin-t distribution.
\clearpage
\section*{Figures}

\clearpage
\section*{Address}
\begin{quote}
Prof Rose Baker, School of Business, University of Salford\\
email rose.baker@cantab.net
\end{quote}
\end{document}